\documentclass[aps,english,prl,twocolumn]{revtex4}
\usepackage{mathrsfs}
\usepackage{amsmath}
\usepackage{graphicx}
\usepackage{amssymb}
\usepackage[dvips]{color}
\usepackage{babel}

\begin{document}

 \title{Non-Magnetic Impurity induced in-gap bound states in two band $s_{\pm}$ superconductors}
\author{  T.K. Ng$^1$ and Y. Avishai$^{1,2}$}
\email {yshai@bgu.ac.il}
\email {phtai@ust.hk}
\affiliation{$^1$ Department of Physics, Hong Kong University of Science and Technology, Clear Water Bay, Kowloon, Hong Kong\\
$^2$ Department of Physics and Ilse Katz Institute for
Nanotechnology, Ben Gurion University of the Negev, Beer Sheva,
Israel} \date{\today }
 \pacs{71.10.Fd,74.20,-z,74.25.Jb}

 \begin{abstract}
 In this paper we study the effect of single non-magnetic impurity in two-band $s$-wave superconductors where the two
 $s$-wave order parameters have relative phase $\delta\neq0$ between them. We show that in-gap bound states are always induced
 by non-magnetic impurities when $\delta=\pi$ ($s_{\pm}$-wave superconductors). The bound state is a consequence of the
 topological nature of the corresponding Bogoliubov-de Gennes equation.
 \end{abstract}

\maketitle

  With the discovery of the Iron-based (pnictides) superconductors, the superconductivity characterized by more than one
 order parameters, i.e. the multi-gap superconductors, becomes a hot topic. Band structure calculations indicate that the
 material has a quasi-two-dimensional electronic structure, with four bands centered around the $\Gamma$- and $M$- points
 in the Brillouin zone contributing to the Fermi surface.  It has been proposed that the superconducting order parameters
 in this multi-band materials have $s$-wave symmetry, but with opposite sign between bands centered at $\Gamma$- and
 $M$-points\cite{hu,mazin,wang}.

  The effect of impurities in this class of materials have been an issue of interests. NMR\cite{NMR1} and lower critical
  field data\cite{field} seems to indicate the existence of nodes in the superconducting order parameter while the
  observations in Angle-resolved photoelectron spectroscopy\cite{ARPES1,ARPES2} favor node-less gaps. One possible
  solution to this controversy is that the order-parameters are $s$-wave-like, but large number of in gap states are
  induced in the material because of the frustrated sign of order-parameters between the bands. Indeed, such a scenario
  has received supports from self-consistent-Born type calculations where in gap states are found to appear
  easily\cite{im1,im2,im3}.

    To have a more precise understanding of the physical effect of non-magnetic impurities and the origin of in-gap states,
  we consider in this paper the effect of a single non-magnetic impurity on two-band $s$-wave superconductors. The
  electrons in the two band are coupled through a Josephson effect which determines the relative sign between the order
  parameters\cite{Ng}. The effect of impurities is studied by analyzing the corresponding Bardeen-Schrieffer-Cooper (BCS)
  theory. In particular we want to understand (1) how the superconducting gap is affected by the impurity and (2) the
  conditions under which in-gapped states are formed. We note that the problem of single impurity has also been tackled
  numerically in particular tight-binding models used to simulate iron pnictides\cite{t1,t2,jhu2}. Our analytical work
  here is independent of the microscopic details of the models and provides results that are complementary to the above
  numerical works.

  \subsection{formulation}
    We start with the Path integral formulation of BCS theory. The system we consider is characterized
    by the BCS action, $S=S_0+S_I$, where
   \begin{equation}
   \label{h0}
     S_0=-\sum_{i,k}\Psi^+_{i}(k)\left(\begin{array}{cc}
     (i\omega_n-\epsilon_{i\vec{k}}+\mu & \Delta_{0i} \\
     \Delta_{0i} & i\omega_n+\epsilon_{i\vec{k}}-\mu \end{array}
     \right)\Psi_{j}(k).
   \end{equation}
   $\Psi_i(k)=\begin{bmatrix}
     c_{i\uparrow}(k) \\ c^+_{i\downarrow}(-k) \end{bmatrix}$,
   $i=1,2$ is the band index and $k=(\vec{k},i\omega_n)$. $\epsilon_{i\vec{k}}$ is the energy dispersion for
   electrons in band $i$ and $c_{\sigma},c^+_{\sigma}$ are spin-$\sigma$ electron annihilation(creation) operators. $S_0$ is
   a sum of two bulk BCS mean-field actions describing two superconducting bands coupled only by Josephson interaction.
   $\Delta_{0i}$ is the superconducting gap when impurities are absent. The effect of a single non-magnetic impurity is
   represented by $S_I$, where
  \begin{equation}
  \label{si}
  S_I={1\over \Omega}\sum_{i,j=1,2,i\omega_n}\Psi_{i}^{\dagger}({i\omega_n})
     \left(\begin{array}{cc}
         U_{ij} & \tilde{\Delta}_{ij} \\
       \tilde{\Delta}_{ij}^* & -U_{ij} \end{array}\right)
      \Psi_{j}(i\omega_n),
  \end{equation}
  where $\Omega=$ volume of system and $\Psi_{i}(i\omega_n)=\sum_{\vec{k}}\Psi_i(k)$.
  $S_I$ describes the effects of an impurity scattering potential $U(\vec{r})\sim\delta^d(\vec{r})
  \sum_{(i,j=1,2),\sigma}U_{ij}c^{(i)+}_{\sigma}(\vec{r})c^{(j)}_{\sigma}(\vec{r})$,
  where $U_{ij}$'s are the scattering matrix element between bands $i$ and $j$ and $\tilde{\Delta}_{ij}=
  \delta_{ij}\tilde{\Delta}_i$ is the induced change in local superconducting gap as a result of the impurity scattering
  potential. We have approximated the induced change in gap to be of form $\tilde{\Delta}_{i}(\vec{r})
  \sim\delta^d(\vec{r})\tilde{\Delta}_{i}$ here, consistent with our simplified form of impurity scattering potential.

  The superconducting order parameters are determined by the mean-field equation
  \begin{equation}
  \label{gap}
  \Delta_{i}(\vec{r})=\Delta_{0i}+\tilde{\Delta}_i(\vec{r})=-\sum_{j=1,2}V_{ij}\langle
  c^{(j)}_{\uparrow}(\vec{r})c^{(j)}_{\downarrow}(\vec{r})\rangle,
  \end{equation}
  where $\langle c^{(j)}_{\uparrow}(\vec{r})c^{(j)}_{\downarrow}(\vec{r})\rangle$ is the pairing amplitude between
  electrons in $j^{th}$ band, $V_{ii}$ represents the pairing interaction between electrons in band $i$ and $V_{12}=
  V_{21}$ is the Josepshon coupling between the pairing order parameters in the two bands.
  %$\tilde{\Delta}_i$ is determined by the mean-field equation\ (\ref{gap}) with impurity included.

  The fermion fields in $S$ can be integrated out to obtain an effective action $S_{eff}$ in terms of $U_{ij}$ and
  $\tilde{\Delta}_{i}$, we obtain
  \[
    S_{eff}=\ln\det(M_0)+\ln\det(1+G_0M_1(U,\tilde{\Delta})),
  \]
  where $\ln\det(M_0)$ is coming from the the mean-field BCS action in the absence of
  impurity and
  \begin{subequations}
  \label{g0}
  \begin{equation}
   G_0(i\omega_n)=M_0(i\omega_n)^{-1}=\left(\begin{array}{cc}
   g_{01}(i\omega_n) & 0 \\
    0 & g_{02}(i\omega_n) \end{array}\right)
  \end{equation}
  where
  \begin{equation}
  g_{0i}(i\omega_n)={\pi N_i(0)\over\sqrt{|\Delta_{0i}|^2-(i\omega_n)^2}}\left(\begin{array}{cc}
  -i\omega_n & \Delta_{0i} \\
  \Delta_{0i}^* & -i\omega_n
  \end{array}\right)\theta(\omega_D-|\omega_n|)
  \end{equation}
  \end{subequations}
  is the (on-site) Nambu matrix Green's function for band $i$ electrons in the absence of the impurity.
  $\omega_D>>|\Delta_{01}|,|\Delta_{02}|$ is the cutoff energy
  for attractive interactions ($\sim$ Debye frequency for phonon superconductors) and $N_i(0)\sim E_F^{-1}$ is the
  band $i$ density of states at the Fermi level. We have assumed $E_F>>\omega_D,\Delta_{01(2)},U_{ij},\omega_n$, etc.
  to justify using a constant density of states in Eq.\ (\ref{g0}). We also have
  \begin{subequations}
  \label{u}
  \begin{equation}
   M_1=\left(\begin{array}{cc}
   W_{11} & W_{12} \\
    W_{21} & W_{22} \end{array}\right),
  \quad~
  W_{ij}=\left(\begin{array}{cc}
  U_{ij} & \delta_{ij}\tilde{\Delta}_{i} \\
  \delta_{ij}\tilde{\Delta}^*_{i} & -U_{ij}
  \end{array}\right).
  \end{equation}
  \end{subequations}
    The free energy associated with the impurity is
  \begin{eqnarray}
  \label{free0}
  F_I & = & -{1\over\beta}\sum_{|\omega_n|<\omega_D}\ln\det(1+G_0M_1(U,\tilde{\Delta}))  \\ \nonumber
     & & +\sum_{ij}(\Delta_{0i}^*+\tilde{\Delta}^*_i)(V^{-1})_{ij}(\Delta_{0j}+\tilde{\Delta}_j)
  \end{eqnarray}
   and the mean-field equation for $\tilde{\Delta}_i$ can be obtained by minimizing the free energy with respect
   to $\tilde{\Delta}_i$. We obtain
  \begin{equation}
  \label{mf0}
  {1\over\beta}\sum_{|\omega_n|,j}V_{ij}\left[(1+G_0(i\omega_n)M_1)^{-1}G_0(i\omega_n)\right]_{j_1,j_2}=
  \tilde{\Delta}_{i}+\Delta_{0i}
  \end{equation}
  where $(j_1,j_2)=(2j-1,2j)$ for $j=1,2$.

  \subsection{single-band case}
     It is helpful to first consider the situation of single band superconductor. In this case the mean-field equation for
   $\tilde{\Delta}$ is
  \begin{equation}
  \label{mf1}
  V'{1\over\beta}\sum_{|\omega_n|<\omega_D}{\pi N(0)(\tilde{\Delta}'+{\Delta_0\over\sqrt{|\Delta_0|^2-(i\omega_n)^2}})\over
  1+|U'|^2+|\tilde{\Delta}'|^2+{2Re(\Delta_0^*\tilde{\Delta}')\over\sqrt{|\Delta_0|^2-(i\omega_n)^2}}}=\Delta_0'+
  \tilde{\Delta}'
  \end{equation}
  where $X'=\pi N(0)X$, where $X=U,V,\tilde{\Delta},\Delta_{0},\omega_D$. The BCS mean-field
  equation in the absence of impurity is recovered if we set $U'=\tilde{\Delta}'=0$. Eq.\ (\ref{mf1}) can be solved
  analytically  in the limit $V',|U'|,|\Delta_0|$ and $\omega_D'=\pi N(0)\omega_D<<1$, which is the case for weakly-coupled
  BCS superconductors. In this case it is straightforward to show that
   \[ \tilde{\Delta}\sim-\Delta_0|U'|^2+O((U',V',\omega_D')^4), \]
    and the effect of impurity is to reduce the gap amplitude at the impurity site. Correspondingly a bound state is
   induced at the impurity site which is determined by the equation $1+G_0(\omega)M_1=0$, or
  \[
  \sqrt{|\Delta_0|^2-\omega^2}(1+U'|^2+|\tilde{\Delta}'|^2)+2Re(\Delta_0^*\tilde{\Delta}')=0.
  \]
   We see that a solution $\omega<|\Delta_0|$ exists when $\Delta_0^*\tilde{\Delta}<0$.
   The solution has energy $\omega\sim\Delta_0(1-2|\Delta_0'|^2|U'|^4)>>\Delta_0-|\tilde{\Delta}|$
   in the limit $|U'|,V',\omega_D'<<1$. The bound state solution is a direct consequence of local suppression of
   superconducting order-parameter by the impurity which creates a local "potential well" in the
   system. The bound state has energy $\omega>$ local gap magnitude $=\Delta_0-|\tilde{\Delta}|$ and is therefore not a true
   ``in-gap" state.

 \subsection{two-band situation}
    Next we consider the two-band situation. To see the new physics associated with the appearance of multiple bands we
 first consider bound states assuming $\tilde{\Delta}_i=0$, i.e. there is no induced local change in the gap amplitudes.

     In this case we obtain after some algebra
   \begin{eqnarray}
   \label{bs2a}
   & & \det(1+G_0(\omega)M_1)=1+{|U_{12}'|^2(|U_{12}'|^2-2U_{11}'U_{22}')\over(1+|U_{11}'|^2)(1+|U_{22}'|^2)}
   \\ \nonumber
   & & -{2\left(\omega^2-|\Delta_{01}||\Delta_{02}|\cos\delta\right)|U_{12}'|^2\over(1+|U_{11}'|^2)(1+|U_{22}'|^2)\sqrt{
   (|\Delta_{01}|^2-\omega^2)(|\Delta_{02}|^2-\omega^2)}}
   \end{eqnarray}
    where $U_{ij}'=\pi\sqrt{N_i(0)N_j(0)}U_{ij}$ and $\delta$ is the relative phase between the two order parameters
    $\Delta_{01}$ and $\Delta_{02}$. The solutions to the equation $\det(1+G_0(\omega)M_2)=0$ can be
    obtained easily since the equation is quadratic in $\omega^2$. We are interested at the bound state solution with
    $\omega<min(|\Delta_{01}|,|\Delta_{02}|)$. Notice that the equation has no solution when $U_{12}=0$, consistent
    with what we observe in the single-band case.

      Assuming that $|\Delta_{02}|>|\Delta_{01}|$, it is easy to see from Eq.\ (\ref{bs2a}) that bound state
    solution with energy $\omega<|\Delta_{01}|$ exists only when $|\Delta_{01}|>|\Delta_{02}|\cos\delta$. In
   particular, no bound state solution exists when $\delta=0$. However bound state exists when $\delta=\pi$, when the two
   superconducting order parameters are out of phase, even when there is no induced local changes in the order parameters.
   % Notice that this is believed to be the relevant case for Iron-based superconductors.
     Solving the equation we obtain
    \begin{eqnarray}
    \label{sol2}
     & &
     \omega^2={1\over2(1-4r^2)}\left(|\Delta_{01}|^{2}+|\Delta_{02}|^{2}+8r^2|\Delta_{01}||\Delta_{02}|\right.
     \\ \nonumber
     & & \left.-(|\Delta_{01}|+|\Delta_{02}|)\sqrt{(|\Delta_{02}|-|\Delta_{01}|)^2+16r^2|\Delta_{01}|
      |\Delta_{02}|}\right)
    \end{eqnarray}
   where
   \[      r={|U_{12}'|^2\over(1+|U_{11}'|^2)(1+|U_{22}'|^2)-|U_{12}'|^2(|U_{12}'|^2-2U_{11}'U_{22}')}.
      \]

    We note that bound state solutions always exist in the limit of small $|U_{12}'|^2$ (Born limit). The solution has
    energy
    \[
    \omega-|\Delta_{01}|\sim-2r^2{(|\Delta_{01}|+|\Delta_{02}|)\over(|\Delta_{02}|-|\Delta_{01}|)}|
    |\Delta_{01}|+O(r^4),
    \]
    in the limit $|\Delta_{02}|-|\Delta_{01}|>>4r\sqrt{|\Delta_{01}||\Delta_{02}|}$ and becomes
    $\omega-|\Delta_{01}|\sim-2|\Delta_{01}|r$
    in the opposite limit $|\Delta_{02}|-|\Delta_{01}|<<4r\sqrt{|\Delta_{01}||\Delta_{02}|}$. More generally,
    it is straightforward to show that solutions with $\omega\geq0$ exists when $r^2\leq1/4$ and no solution exists at
    $r^2>1/4$. Therefore there is an intermediate range of parameters $U$'s where bound state solution does not exist. At
    around the resonance point $r^2=1/4-\delta$ we obtain
    \[
    \omega^2=4\delta{|\Delta_{01}|^2|\Delta_{01}|^2\over|\Delta_{01}|^2+|\Delta_{02}|^2}+O(\delta^2).
    \]

      Our result indicates that the existence of in-gap state and the corresponding bound state energy depends on the
    particular form of impurity scattering potential which determines the parameters $U_{11},U_{22}$ and
    $U_{12}$. Assuming the $U$'s are all proportional to each other we find that in the strong scattering (Unitary) limit
    $|U_{ij}'|^2\rightarrow\infty$, $r^2\rightarrow0$ and the bound state energy approaches zero asymptotically.
    This result is in qualitative agreement with tight-binding calculations\cite{t1,t2,jhu2} where shallow in-gap
    states are found to exist easily in two-band superconductors with order parameters of opposite
    sign. Our non-model-specific result suggests that $\omega\rightarrow0$ bound states are in general
    allowed in two band superconductors with frustrated sign between order parameters.

    To examine whether the in-gap state is robust against changes in the  superconducting order parameters we consider the
    case of symmetric bands with $g_{01}(i\omega)=g_{02}(i\omega)$ and $\Delta_{0i}=\Delta_0e^{i\phi_i}$, i.e. the two bands
    differ only in the phase of the order parameters. To simplify the problem further we
    set $U_{11}=U_{22}=0$ so that
    \begin{equation} \label{MatrixM1}
    M_1=\begin{pmatrix} 0&\tilde{\Delta}e^{i \theta_1}& U &0 \\ \tilde{\Delta}e^{-i \theta_1}&0&0&-U \\
    U&0&0& \tilde{\Delta}e^{i \theta_2}\\ 0&-U& \tilde{\Delta}e^{-i \theta_2}&0 \end{pmatrix},
    \end{equation}
   where $\tilde{\Delta}$ and $\theta_i$ are to be solved self-consistently from the mean-field equation\ (\ref{mf0}).
   The determinant $\det(1+G_0M_1)$ can still be computed analytically in this case. We obtain after lengthy algebra
   \begin{eqnarray}
   \label{det2}
   & & \det(1+G_0(i\omega)M_1)=A(i\omega)+B(i\omega)[\cos(\theta_1-\phi_1)
   \\ \nonumber
   & & +\cos(\theta_2-\phi_2)]+C(i\omega)[\cos(\theta_2-\phi_1)+\cos(\theta_1-\phi_2)]    \\ \nonumber
   & &+D(i\omega)\cos(\phi_1-\phi_2)+2|U_{12}'|^2|\tilde{\Delta}'|^2\cos(\theta_1-\theta_2)
   \\ \nonumber
   & &+E(i\omega)\cos(\theta_1-\phi_1)\cos(\theta_2-\phi_2),
   \end{eqnarray}
   where
  \[
  A(i\omega)=(1+|\tilde{\Delta}'|)^2+|U'|^4+2{(i\omega)^2|U'|^2\over(i\omega)^2-|\Delta_0|^2},
  \]
  \[
  B(i\omega)=-{2\Delta_0\tilde{\Delta}'\sqrt{|\Delta_0|^2-(i\omega)^2}[1+|\tilde{\Delta}'|^2]\over(i\omega)^2-
  |\Delta_0|^2}, \]
 \[ C(i\omega)=-{2 \Delta_0\tilde{\Delta}'
 \sqrt{|\Delta_0|^2-(i\omega)^2}|U'|^2\over(i\omega)^2-|\Delta_0|^2},
 \]
  \[
  D(i\omega)=-{2|\Delta_0|^2|U'|^2\over(i\omega)^2-|\Delta_0|^2}, E(i\omega)=
  -{2|\Delta_0|^2|\tilde{\Delta}'|^2\over(i\omega)^2-|\Delta_0|^2},
  \]
 where $X'=\pi N(0)X$ as before.

   The mean-field equation can be solved in the weak impurity scattering limit $|U'|,|\Delta_0'|,|V_{ij}'|,\omega_D'<<1$.
   Keeping only terms to order $|U'|^2$ and $\tilde{\Delta}'$ in the mean-field equation, we obtain
   $\theta_i=\phi_i (i=1,2)$ and
   \[
   \tilde{\Delta}\sim|U'|^2\Delta_0\cos(\phi_1-\phi_2). \]
   Notice that the superconducting order parameter is {\em enhanced} by scattering between the two bands if
   $\phi_1=\phi_2$ but is {\em suppressed} by scattering if $\phi_1-\phi_2=\pm\pi$, suggesting that non-magnetic
   impurity induces a local {\em ferromagnetic} Josephson coupling between the superconducting
   order parameters which disfavors a $s_{\pm}$ state\cite{im2}.

    We next examine the solution(s) to the equation $\det(1+G_0(\omega)M_1)=0$ with $\omega^2<|\Delta_0|^2$. Defining
  $y^2 \equiv\Delta_0^2-\omega^2$
  %the expression for $\det(1+G_0(\omega)M_1)$ becomes a second degree polynomial in $y$ and
   we find that a zero at some $0 < y_0 <\Delta_0$ implies an in-gap bound state. Solving the equation (with $\theta_i=\phi_i
  (i=1,2)$) we obtain $ y_0=(-\beta+\gamma)/\alpha$,  where
  \[ \beta=2 \Delta_0 \tilde{\Delta}' \{ 1+|\tilde{\Delta}'|^2+|U'|^2 \cos(\phi_1-\phi_2),  \]
 \[  \alpha=1+2 (|U'|^2+|\tilde{\Delta}'|^2)+[U'^4+|\tilde{\Delta}'|^4+2|U'|^2|\tilde{\Delta}'|^2
 \cos(\phi_1-\phi_2)],  \]
 and
 \[
 \gamma^2=2|U'|^2\Delta_0^2[1+|U'|^2+|\tilde{\Delta}'|^2]^2[1-\cos(\phi_1-\phi_2)].  \]
   It is easy to see that
   \begin{subequations}
   \begin{equation}
   \label{y1}
     y_0=-{2\Delta_0\tilde{\Delta}'\over1+|U'|^2+|\tilde{\Delta}'|^2}
   \end{equation}
   for $\phi_1=\phi_2$ and a $y>0$ solution does not exist for small $|U_{12}'|^2$ where $\Delta_0 \tilde{\Delta}'>0$.
    The situation is very different for $\phi_1-\phi_2=\pm\pi$. In this case
   \begin{eqnarray}
   \label{y2}
   y_0 & = & 2\Delta_0.{|U'|(|(1+|\tilde{\Delta}'|^2+|U'|^2)-\tilde{\Delta'}(1+|\tilde{\Delta}'|^2-|U'|^2)
   \over1+2(|\tilde{\Delta}'|^2+|U'|^2)+(|\tilde{\Delta}'|^2-|U'|^2)^2} \\ \nonumber
   & & \sim2\Delta_0|U'|
   \end{eqnarray}
   and $\omega\sim\Delta_0-2|U'|^2\Delta_0$
   \end{subequations}
   in the limit $|U'|,|\tilde{\Delta}'|<<1$. Notice that the bound state has energy below the ``local" gap magnitude
   $\Delta_0+\tilde{\Delta}\sim\Delta_0(1-|U'|^2)$, indicating that the formation of in-gap bound
   state is robust to local gap distortion.

 \subsection{Impurity averaged single particle density of states}
\begin{figure}[!h]
 \centering
 \includegraphics[width=5truecm, angle=90]{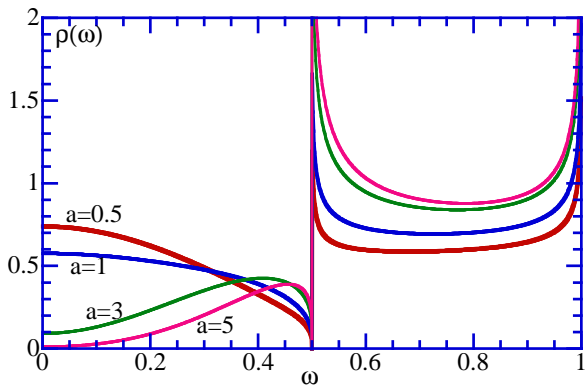}
 \caption{\footnotesize{The averaged density of states $\rho(\omega)$ for $a=0.5,1,3, 5$ and $\alpha=0.4$ with $|\Delta_{02}'|=0.5,
 |\Delta_{01}'|=1$ and $N_1(0)=N_2(0)$.}} \label{Fig1}
 \end{figure}

\begin{figure}[!h]
\centering
\includegraphics[width=5truecm, angle=90]{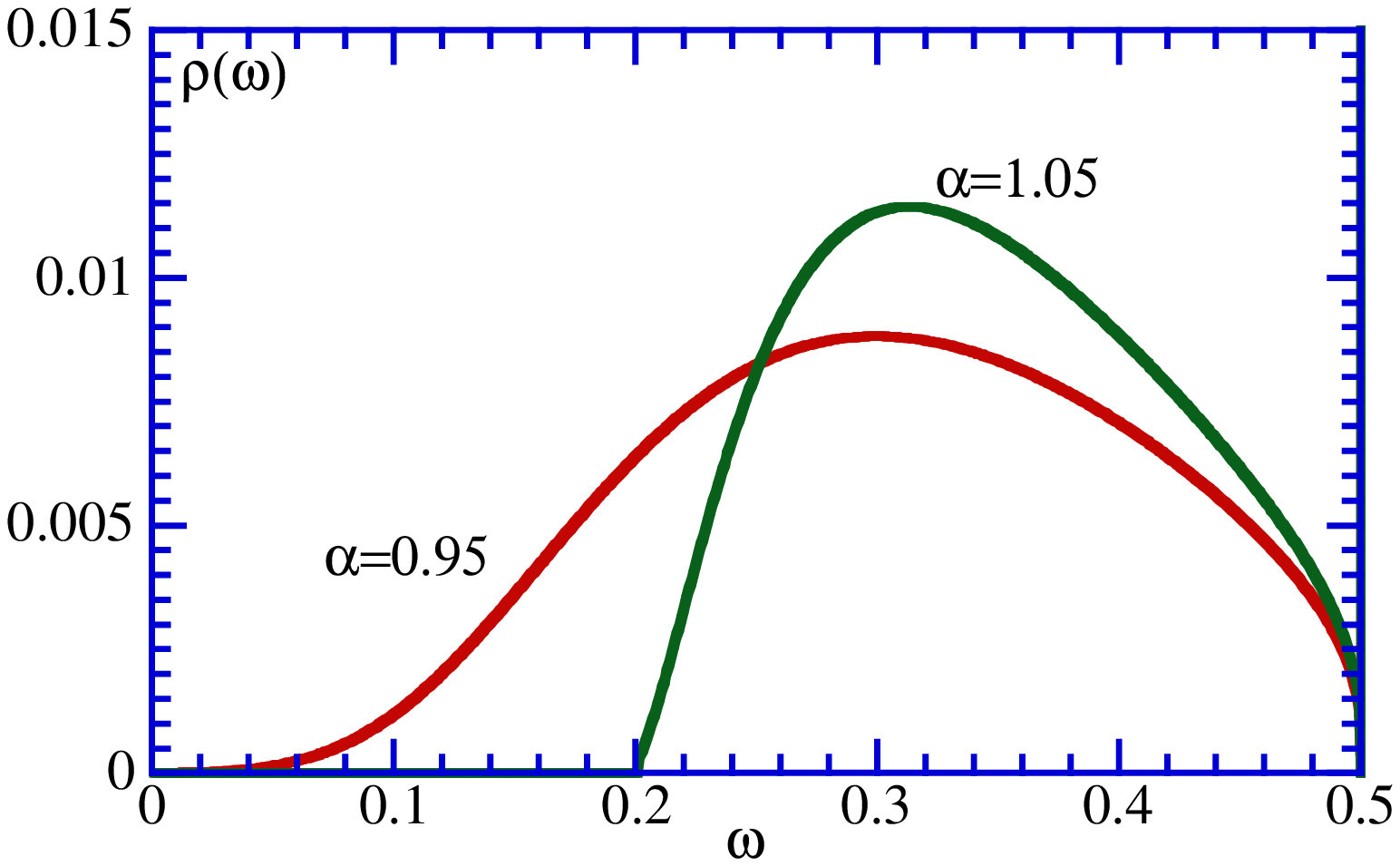}
 \caption{\footnotesize{The averaged density of states $\rho(\omega)$ for $a=1$ and $\alpha=0.95, 1.05$ with $|\Delta_{02}'|=0.5,
 |\Delta_{02}'|=1$ and $N_1(0)=N_2(0)$.}} \label{Fig2}
 \end{figure}

   To have a more quantitative feeling of the effect of impurities we compute the single-particle density of states in
   our model with $U_{11}=U_{22}=\alpha U_{12}=\alpha U$ and $\tilde{\Delta}=0$. In this case the matrix Green's function
   $G(\omega;U)=(1+G_0(\omega)M_1(U))^{-1}G_0(\omega)$ can be evaluated exactly and the impurity averaged Green's
   function
   \begin{equation}
   \label{imav}
     \langle G(\omega)\rangle=\int dUG(\omega;U)P(U)
   \end{equation}
   can be evaluated for given impurity potential distribution $P(U)$. Contrary to self-consistent Born type calculations
   the calculation here is valid only in the limit of low concentration of impurities where interference effects between
   different impurity scattering events are negligible. We find that $\langle G_{12}(\omega)\rangle=\langle G_{21}
   (\omega)\rangle=0$ for even distribution $P(U)=P(-U)$ and only intra-band Green's function
   survives impurity average. (Inter-band Green's function will contribute when evaluating two-particle
   correlation functions.) The trace of the total electron Green's function is
 \begin{eqnarray}
 && \mbox{Tr} G(\omega)=-2 \Gamma \omega
 [1+U'^2(1+\alpha^2)] \nonumber \\
 &&  \frac {S_1(\omega) + S_2(\omega) ]}
 {a(\omega) U'^4+b(\omega) U'^2+c(\omega)},
 \label{NewTraceG}
 \end{eqnarray}
 with
 \begin{eqnarray}
 && a(\omega)=(1-\alpha^2)^2S_1(\omega)S_2(\omega), \nonumber \\
 && b(\omega)=-2\left [ \omega^2-|\Delta_{01} \Delta_{02}| \cos \delta
 -\alpha^2 S_1(\omega) S_2 (\omega) \right ], \nonumber \\
 && c(\omega)= S_1(\omega) S_2(\omega).
 \label{abc}
 \end{eqnarray}
   The single-particle density of states given by $\rho(\omega)=\Im\mbox{Tr} G(\omega)$ is evaluated numerically for
   $s_{\pm}$ superconductors ($\delta=\pi$) with $P(U)=\sqrt{\frac{a}{\pi}} e^{-aU'^2}$ for two different values of
   $a=0.5$ to $3$ and $\alpha=0.4$ with $|\Delta_{02}'|=0.5, |\Delta_{01}'|=1$ and $N_1(0)=N_2(0)$. The results are shown
   in fig.(\ref{Fig1}).
  The density of states for $\alpha=0.95, 1.05$ and $0 \le \omega \le \Delta_{02}$
   is also shown in fig.(\ref{Fig2}) for comparison.
 We find that nonzero density of state $\rho(\omega))$ is induced inside the gap with nonzero spectral weight on the Fermi
 surface in general. The precise form of $\rho(\omega)$ depends also on $U_{11},U_{22}$ and the distribution of impurity
 scattering strength $P(U)$. The in-gap spectral weight increases with decreasing $a$ and shifts to lower energy with
 decreasing $\alpha$, indicating that in-gap states are strengthened by strong inter-band scattering but suppressed by
 intra-band scattering. Notice also that for $\alpha>1$, $r<1/2$ and no $\omega=0$ bound state exists! The in-gap states
 have energy $\omega>\omega_c\sim0.2$ for $\alpha=1.05$ as shown in fig.(\ref{Fig2}). We cautioned that we haven't included the induced
 changes in gap functions $\tilde{\Delta}$'s in our calculation. Thus we expect that the near gap edge behavior in our
 calculation may not be reliable but the deep in-gap behavior should be qualitatively correct.

  Our result indicates that non-magnetic impurities is a relevant perturbation to the physics of multi-band superconductors
  with frustrated sign between order parameters. They introduce a ``ferromagnetic" Josephson coupling between the two
  superconducting order parameters which disfavors the $s_{\pm}$ state and introduce in-gap bound states in the quasi-particle
  spectrum. The energy spectrum of the in-gap states depend on the nature of the
  impurity potential and is ``non-universal".

 \subsection{Origin of the in-gap states}

   It is important to understand why in-gap bound states exist so easily when the relative phase between the two
   superconducting order parameters is $\pi$, independent of the microscopic details of the system. The
   robustness of the in-gap bound states can be understood if we notice that the approximate mean-field Green's functions\
   (\ref{g0}) we used in our calculation is the exact Green's function of a corresponding one-dimensional superconductor
   problem, if we linearize the fermion spectrum $\epsilon_{i\vec{k}}-\mu\rightarrow\pm v_{fi}(k-k_{Fi})$ in action\
   (\ref{h0}). In this case the density of state $N_i(\epsilon)\sim dk/d\epsilon$ becomes constant and Eq.\ (\ref{g0})
   becomes exact. In this representation, the impurity introduces finite tunnelling probability between two
   one-dimensional superconductors, one located on the left of the impurity, and the other one on the
   right when $U_{12}\neq0$. This problem has been studied in
   Refs.\cite{Ng} and \cite{feng}, where bound states are found to exist when the phase difference between the two
   superconductors is $\delta=\pi$.

     The existence of in-gap bound state can be understood by noting that the one-dimensional Bogoliubov equation
     with a linearized electron spectrum around the Fermi surface is essentially a Dirac equation for spinless fermions at
     one-dimension\cite{Ng}. In particular, the gap function $\Delta$ becomes the mass term in the Dirac fermion
     representation. Thus the problem is mathematically equivalent to a tunnelling problem between two species of Dirac
     fermions with masses of opposite sign. For perfect tunnelling, it is known that a $\omega=0$ mid-gap state exists if the
     masses of the Dirac fermions have opposite sign at the two sides of the tunnelling barrier because of the topological
     structure of the problem\cite{atiyah}. The bound states split into two with energies $\pm\omega>0$ when a tunnelling
     barrier exists\cite{feng,jhu} and eventually merge into the continuum spectrum when the tunnelling barrier is high
     enough. This is in qualitative agreement with what we find here when $\tilde{\Delta}_i=0$. In
     addition, we also observed that the induced $\tilde{\Delta}_i$ is small and does not affect the in-gap bound states
     in the limit $|U_{12}'|,|\Delta_0'|,|V_{ij}'|<<1$ in the case $|\Delta_{01}|=|\Delta_{02}|$ and $U_{11}=U_{22}=0$.

       Lastly, we comment that the effect of magnetic impurity can also be included straightforwardly by introducing a
     magnetic scattering term $V_{ij}^{\sigma\sigma'}(\vec{r})\sim\delta(\vec{r})\sigma V_{ij}\delta_{\sigma\sigma'}$ in
     our calculation. It is known that magnetic scattering introduces in-gap bound states even for single-band
     superconductors and this feature remains for two-band superconductors without phase frustration. Therefore magnetic
     impurities do not induce qualitative differences between two-band $s$-wave and $s_{\pm}$-wave superconductors\cite{jhu2}
     which is why we concentrate only on non-magnetic impurities in the present paper.

    Summarizing, by studying the one-impurity problem carefully, we show in this paper that (non-magnetic) impurity is a
  relevant perturbation in two-band $s_{\pm}$-wave superconductors where the order parameters are of opposite signs. The
  suppression of $s_{\pm}$ state and generation of in-gap bound states in quasi-particle spectrum are natural phenomena
  associated with non-magnetic impurities and should be taken into
  account carefully in understanding the electronic properties of Iron-base superconductors.\\
  \ \\
  {\bf Acknowledgment} T.K.N acknowledges support by HKUGC through grant CA05/06.Sc04.
  The research of Y.A is partially supported by an ISF grant.

\end{document}